\documentclass[12pt,thmsa]{article}
\usepackage{sw20lart}

\input{tcilatex}
\begin{document}

\author{Emilio Santos, Emeritus Professor of Theoretical Physics \and Universidad de
Cantabria. Santander. Spain}
\title{The quantum electromagnetic field in the Weyl-Wigner representation}
\date{December, 3, 2024}
\maketitle

\begin{abstract}
The quantum electromagnetic (EM) field is formulated in the Weyl-Wigner
representation (WW), which is equivalent to the standard Hilbert space one
(HS). In principle it is possible to interpret within WW all experiments
involving the EM field interacting with macroscopic bodies, the latter
treated classically. In the WW formalism the essential difference between
classical electrodynamics and the quantum theory of the EM field is\ just
the assumption that there is a random EM field filling space\textit{, }i.e.
the existence of a zero-point field with a Gaussian distribution for the
field amplitudes. I analyze a typical optical test of a Bell inequality. The
model admits an interpretation compatible with local realism, modulo a
number of assumptions assumed plausible.

\end{abstract}

\section{Introduction}

\subsection{The standard quantization method for fields}

Quantum field theory started with the electromagnetic field $\left(
EM\right) $ quantization by Dirac in 1927. As is well known the procedure
consists of expanding the field in plane waves, or more generally normal
modes, writing every mode in terms of two complex conjugated time-dependent
amplitudes (c-numbers) $\left\{ a_{j}\left( t\right) ,a_{j}^{*}\left(
t\right) \right\} ,$ where $j$ labels a mode, and then promoting the
amplitudes to be operators in a Hilbert space \cite{Milonni}.

In order to define the commutation properties of the operators it is
convenient to introduce two auxiliary quantities $\left\{ x_{j}\left(
t\right) ,p_{j}\left( t\right) \right\} $ as follows 
\begin{equation}
x_{j}\left( t\right) \equiv \frac{c}{\sqrt{2}\omega _{j}}\left( a_{j}\left(
t\right) +a_{j}^{*}\left( t\right) \right) ,p_{j}\left( t\right) \equiv 
\frac{i
\rlap{\protect\rule[1.1ex]{.325em}{.1ex}}h%
\omega _{j}}{\sqrt{2}c}\left( a_{j}\left( t\right) -a_{j}^{*}\left( t\right)
\right) ,  \label{10}
\end{equation}
where $i$ is the imaginary unit. Then the (Maxwell) evolution of every field
amplitude $a_{j}\left( t\right) $ corresponds to the time change of the
quantities ($x_{j}\left( t\right) ,p_{j}\left( t\right) )$ as if they were
the coordinate and momentum of a harmonic oscillator with unit mass and
proper frequency $\omega _{j}.$ Therefore the EM field may be treated
formally as a collection of harmonic oscillators. (For the sake of clarity I
write operators in a Hilbert space with a `hat', e. g. $\hat{a}_{j},\hat{a}%
_{j}^{\dagger },$ and numerical (c-number) amplitudes without `hat', e. g. $%
a_{j},a_{j}^{*}).$ Then the field is quantized via the said oscillators, as
in elementary quantum mechanics, promoting $\left\{ x_{j}\left( t\right)
,p_{j}\left( t\right) \right\} $ to be operators fulfilling the standard
commutation rules. Hence field quantization is completed with the following
commutation rules at equal times 
\begin{equation}
\left[ \hat{a}_{j},\hat{a}_{k}\right] =\left[ \hat{a}_{j}^{\dagger },\hat{a}%
_{k}^{^{\dagger }}\right] =0,\left[ \hat{a}_{j},\hat{a}_{k}^{\dagger
}\right] =\delta _{jk},  \label{com}
\end{equation}
$\delta _{jk}$\ being Kronecker delta. The operators $\hat{a}_{j}$ $\left( 
\hat{a}_{j}^{\dagger }\right) $ are usually named annihilation (creation)
operators of photons.

Dirac\'{}s procedure is the standard quantization method for fields. It
rests on the mathematical theory of Hilbert spaces. In this article I will
study an alternative formalism for the quantum EM field deriving from the
work made by H. Weyl \cite{Weyl} and E. P. Wigner\cite{Wigner} on
(non-relativistic) quantum mechanics of particles (QM\ in the following).

\subsection{Wigner representation in quantum mechanics}

In 1927 Weyl proposed a quantization method for systems of particles via a
transform that converts classical (c-number) coordinates and momenta into
operators in a Hilbert space \cite{Weyl}. The idea may be put as follows. If
we have a classical function in phase space, $F(\mathbf{x,p}),$ involving
the position and momentum of a particle, we may get a quantum counterpart, $%
Q(\mathbf{\hat{x},\hat{p}}),$ involving operators in a Hilbert space via the
following symbolic integral 
\[
Q(\mathbf{\hat{x},\hat{p}})=\int d\mathbf{x}\delta \left( \mathbf{x-\hat{x}}%
\right) \int d\mathbf{p}\delta \left( \mathbf{p-\hat{p}}\right) F(\mathbf{x,p%
}), 
\]
where $\delta $ is Dirac delta. The integrals extend over the whole 3D
space. The delta function of an operator is not well defined, but we may
give a meaning to the symbolic equation substituting an integral
representation for the ``deltas'', that is 
\[
Q(\mathbf{\hat{x},\hat{p}})=\frac{1}{4\pi ^{2}}\int d\mathbf{\lambda }\int d%
\mathbf{\mu }\int d\mathbf{x}\exp \left[ i\mathbf{\lambda }\cdot \left( 
\mathbf{x-\hat{x}}\right) \right] \int d\mathbf{p}\exp \left[ i\mathbf{\mu }%
\cdot \left( \mathbf{p-\hat{p}}\right) \right] F(\mathbf{x,p}), 
\]
where $\mathbf{\lambda }$ and $\mathbf{\mu }$ are vector variables. This
equation presents a problem, namely that the function $Q$ may change if the
order of the \textbf{x} and \textbf{p} integrals is reversed, due to the
fact that the operators $\mathbf{\hat{x}}$ and $\mathbf{\hat{p}}$ do not
commute. Hence there is an ambiguity in the ordering of the operators and
Weyl chose the following symmetrical order, that is 
\begin{equation}
Q(\mathbf{\hat{x},\hat{p}})=\frac{1}{4\pi ^{2}}\int d\mathbf{\lambda }\int d%
\mathbf{\mu }\int d\mathbf{x}\int d\mathbf{p}\exp \left[ i\mathbf{\lambda }%
\cdot \left( \mathbf{x-\hat{x}}\right) +i\mathbf{\mu }\cdot \left( \mathbf{p-%
\hat{p}}\right) \right] F(\mathbf{x,p}),  \label{00}
\end{equation}
where $Q(\mathbf{\hat{x},\hat{p}})$ is called Weyl transform of $F(\mathbf{%
x,p}).$ The generalization to a system of particles is straightforward. The
resulting equations from eq.$\left( \ref{00}\right) $ do not agree with
those of QM in general, whence Weyl quantization method is not too useful.

In 1932 Wigner \cite{Wigner} proposed a method to achieve the reverse of
quantization via introducing a formalism with classical appearance for
quantum mechanics. He proposed to get a function $W_{\psi }\left( \mathbf{r},%
\mathbf{p,}t\right) $ in the phase space of the particle from the quantum
wavefunction $\psi \left( \mathbf{r,}t\right) $ of a state as follows 
\begin{equation}
W_{\psi }\left( \mathbf{r},\mathbf{p.}t\right) =\int d\mathbf{u}\psi
^{*}\left( \mathbf{r+u,}t\right) \psi \left( \mathbf{r-u,}t\right) \exp
\left( 2i\mathbf{u\cdot p}\right) ,  \label{01}
\end{equation}
where $W_{\psi }\left( \mathbf{r},\mathbf{p}\right) $ is named Wigner
function of the state $\psi \left( \mathbf{r}\right) $. The evolution
equation, in the form of a time derivative of $W_{\psi }\left( \mathbf{r},%
\mathbf{p.}t\right) ,$ may be obtained taking Schr\"{o}dinger equation into
account but I omit the (cumbersome) expression \cite{Scully}, \cite{Zachos}.
It is also possible to get the Wigner function when the state is given by a
density operator in the abstract Hilbert space, rather than a wavefunction
as in eq.$\left( \ref{01}\right) $. The Wigner function of a state $\hat{M}$
is obtained in this case as follows 
\begin{eqnarray}
W_{\hat{M}} &=&T_{W}\left[ \hat{M}\right] \equiv \frac{1}{4\pi ^{2}}\int d%
\mathbf{\lambda }\int d\mathbf{\mu }\exp \left[ -i\mathbf{\lambda }\cdot 
\mathbf{x}-i\mathbf{\mu }\cdot \mathbf{p}\right]  \nonumber \\
&&\times Tr\left\{ \hat{M}\exp \left[ i\mathbf{\lambda }\cdot (\mathbf{x}-%
\mathbf{\hat{x}})+i\mathbf{\mu }\cdot \left( \mathbf{p}-\mathbf{\hat{p}}%
\right) \right] \right\} ,\smallskip \smallskip  \label{02}
\end{eqnarray}
where $Tr\left\{ {}\right\} $ means the trace operation. The Wigner
transform eq.$\left( \ref{02}\right) $ is actually the inverse of the Weyl
transform when the operator $\hat{M}$ may be written as a function of the
canonical operators $\mathbf{\hat{x}}$ and $\mathbf{\hat{p}.}$

The Wigner transform (or inverse Weyl transform) eq.$\left( \ref{02}\right) $
may be applied not only to density operators representing states but also to
observables, with the result that the whole QM may be formulated in terms of
phase space functions. That formalism is usually named \textit{Wigner
representation of QM }\cite{Scully}, \cite{Zachos}. In particular the
Schr\"{o}dinger equation of a system of particles becomes a diferential
equation for the Wigner function $W_{\psi }\left( \left\{ \mathbf{r}_{j},%
\mathbf{p}_{j}\right\} ,t\right) $ and the expectation value of the
observable \^{M} in the state $\psi $ becomes an integral 
\[
\left\langle \psi \left| \hat{M}\right| \psi \right\rangle =\int W_{M}\left(
\left\{ \mathbf{r}_{j},\mathbf{p}_{j}\right\} \right) W\psi \left( \left\{ 
\mathbf{r}_{j},\mathbf{p}_{j}\right\} ,t\right) \prod_{j}d\mathbf{r}_{j}d%
\mathbf{p}_{j}. 
\]
The Wigner representation is equivalent to the standard Hilbert space
formulation of QM and it is useful for some purposes. However it does not
solve the problems of interpretation because it does not admit a realistic
interpretation. In fact the Wigner functions of states are positive definite
but for a slight fraction of pure quantum states, that is when the
wavefunction is Gaussian \cite{Soto}. Therefore they cannot be interpreted
as probability distributions in phase space. Here I name realistic the
interpretation of a formalism fitting in the celebrated EPR article \cite
{EPR}.

The Wigner representation of QM may be extended to the electromagnetic
field. In fact the inverse of the Weyl transform may be written, taking the
change of variables eqs.$\left( \ref{10}\right) $ into account, as follows 
\begin{eqnarray}
W_{\hat{M}} &=&T_{W}\left[ \hat{M}\right] \equiv \frac{1}{(2\pi ^{2})^{n}}%
\prod_{j=1}^{n}\int_{-\infty }^{\infty }d\lambda _{j}\int_{-\infty }^{\infty
}d\mu _{j}\exp \left[ -2i\lambda _{j}\mathrm{Re}a_{j}-2i\mu _{j}\mathrm{Im}%
a_{j}\right]  \nonumber \\
&&\times Tr\left\{ \hat{M}\exp \left[ i\lambda _{j}\left( \hat{a}_{j}+\hat{a}%
_{j}^{\dagger }\right) +\mu _{j}\left( \hat{a}_{j}-\hat{a}_{j}^{\dagger
}\right) \right] \right\} ,\smallskip \smallskip  \label{Wtransform}
\end{eqnarray}
where $T_{W}\left[ \hat{M}\right] $ stands for (inverse) Weyl transform of
the operator $\hat{M}.$ The result obtained, $W_{\hat{M}}\left( \left\{
a_{j},a_{j}^{*}\right\} \right) ,$ is a function of (c-number) field
amplitudes. Here the operator $\hat{M}$ may be either an observable or the
density operator of a state. That is the transform eq.$\left( \ref
{Wtransform}\right) $ obtains the passage from the standard (HS) quantum
formalism to an alternative Weyl-Wigner (WW) representation for the same
theory in terms of the amplitudes $\left\{ a_{j},a_{j}^{*}\right\} $ of the
EM field.

Unlike standard QM, the WW representation of the quantized EM field \textit{%
does suggest a realistic picture for the field} as argued below in this
article. In sections 2 to 4 I will be concerned mainly with the formal
aspects of WW, leaving further discussion about interpretation for the last
section 5. The WW representation of the quantized EM field has been studied
elsewhere \cite{Frontiers}, \cite{FOOP}, \cite{EPJP}, \cite{book} . In this
article I will present further elaboration and applications of the
formalism. A less formal, semi-quantitative, realistic interpretation of the
quantized EM field has been provided elsewhere \cite{Foundations} for many
phenomena currently assumed to be purely quantum.

In section 2 I shall derive the most relevant properties of the WW formalism
got from HS via the inverse Weyl transform eq.$\left( \ref{Wtransform}%
\right) $. After that I will discuss two matters related to the WW
formalism, namely whether it may be seen as a new method for field
quantization in section 3, and the usefulness of the WW formalism for the
interpretation of some experiments in sec. 4. Several aspects of
interpretation will be discussed in section 5.

\section{From Hilbert space to Weyl-Wigner representation of fields}

Here I will study the EM field in the Coulomb gauge because it is more
appropriate than the Lorentz gauge for the applications considered in
section 4, although the latter is more common and useful in (relativistic)
quantum electrodynamics. Maxwell theory is compatible with special
relativity, which is explicitly exhibited in the covariant Lorentz gauge.
However it may be formulated in the Coulomb gauge, in terms of a vector
potential \textbf{A}(\textbf{x},t) plus a scalar potential $\phi $(\textbf{x}%
,t), the latter taking account of the instantaneous electrostatic
interaction between charges (but this does not mean that the theory violates
relativistic invariance) \cite{Sakurai}. Then the quantity expanded in plane
waves is the vector potential. The effect of the electrostatic interaction
is straightforward and will be ignored in the following.

In free space the expansion may be written 
\begin{equation}
\mathbf{A}(\mathbf{x},t)=\sum_{l}a_{l}\mathbf{\varepsilon }_{l}\exp \left( i%
\mathbf{k}_{l}\mathbf{\cdot x-}i\omega _{l}t\right) +\func{complex}conjugate,
\label{A}
\end{equation}
where $\mathbf{k}_{l}$ is the wavevector and $\mathbf{\varepsilon }_{l}$ the
polarization vector of a mode with frequency $\omega _{l}=c\left| \mathbf{k}%
_{l}\right| $. As said abvove the standard quantization of the field
consists of introducing a Hilbert space and promoting the classical
amplitudes to be operators $\left\{ \hat{a}_{j},\hat{a}_{j}^{\dagger
}\right\} $\ acting on that space. At a difference with eq.$\left( \ref{10}%
\right) ,$ here and in the following I will write explicitly the time
dependence so that the operators $\left\{ \hat{a}_{j},\hat{a}_{j}^{\dagger
}\right\} $ are time independent and they fulfil the commutation rules eq.$%
\left( \ref{com}\right) .$

In the following I summarize the most relevant properties of WW.

\textbf{Transform of products of field amplitudes.} Eq.$\left( \ref
{Wtransform}\right) $ allows deriving the product of amplitudes in WW for
any HS product of creation and annihilation operators. If we have a product
of WW amplitudes like $a_{j}^{m}a_{j}^{*n}$ the HS counterpart is 
\begin{equation}
a_{j}^{m}a_{j}^{*n}\rightarrow \left( \hat{a}_{j}^{m}\hat{a}_{j}^{\dagger
n}\right) _{sym},  \label{sym}
\end{equation}
where \textit{sym} stands for symmetric and it means writing a sum of the $%
m+n$ operator products in all possible orderings and then dividing by the
number of terms that is $\left( m+n\right) !/(m!n!)$. The reverse passage
from HS to WW requires a prior rewriting the HS expression as a sum of
products of operators in symmetrical order, taking the commutation rules
into account.

\textbf{States. }In WW the states are defined via the (inverse) Weyl
transform of the HS states. Thus any state in WW, corresponding to a given
state in HS, is a function of the amplitudes $\left\{
a_{j},a_{j}^{*}\right\} ,$ named ``Wigner function'' of the state.

\textbf{The ``vacuum'' state. }In HS the ground state of the free field
(called the vacuum state) is represented by a vector $\mid 0\rangle $\
defined by 
\begin{equation}
\hat{a}_{j}\mid 0\rangle =0\Rightarrow \langle 0\mid \hat{a}_{j}^{\dagger
}=0,\text{ for all radiation modes,}  \label{0}
\end{equation}
$0$\ being here the null vector in the HS. Alternatively it may be
represented by the density operator 
\begin{equation}
\hat{\rho}=\mid 0\rangle \langle 0\mid .  \label{2d}
\end{equation}
If this operator is inserted in place of $\hat{M}$ in eq.$\left( \ref
{Wtransform}\right) $ we get after some algebra the Wigner function of the
vacuum state, that is \cite{FOOP}, \cite{EPJP},

\begin{equation}
W_{0}\left( \left\{ a_{j}\right\} \right) =\prod_{j}\frac{2}{\pi }\exp
\left( -2\left| a_{j}\right| ^{2}\right) ,  \label{1}
\end{equation}
which is normalized for the integration with respect to $\prod_{j}d\func{Re}%
a_{j}d\func{Im}a_{j}.$ Hence the mean square average value of each amplitude
is 
\begin{equation}
\left\langle \left| a_{j}\right| ^{2}\right\rangle =\frac{1}{2}.  \label{1a}
\end{equation}
If eq.$\left( \ref{1}\right) $ was interpreted as a probability distribution
then eq.$\left( \ref{1a}\right) $ would be the variance of $\left|
a_{j}\right| .$ The field represented by eq.$\left( \ref{1}\right) $ will be
labeled the ``zeropoint field'' (ZPF).

Taking the last eq.$\left( \ref{Hamiltonians}\right) $ (see below) into
account, eq.$\left( \ref{1}\right) $ leads to the following distribution of
the energy amongst the radiation modes 
\begin{equation}
W_{E}\left( \left\{ E_{j}\right\} \right) =\prod_{j}\frac{2}{%
\rlap{\protect\rule[1.1ex]{.325em}{.1ex}}h%
\omega _{j}}\exp \left( -\frac{2E_{j}}{
\rlap{\protect\rule[1.1ex]{.325em}{.1ex}}h%
\omega _{j}}\right) ,  \label{W0}
\end{equation}
where $E_{j}$ is the energy of mode $j$ and the normalization is appropriate
for integration with respect to $\prod_{j}dE_{j}$. I point out that eq.$%
\left( \ref{W0}\right) $ leads to the following spectral density (energy of
the field per unit volume and unit frequency interval) 
\begin{equation}
\rho \left( \omega \right) =\frac{1}{2}
\rlap{\protect\rule[1.1ex]{.325em}{.1ex}}h%
\omega ^{3},  \label{spectrum}
\end{equation}
which possesses the important property of being Lorentz invariant. Indeed
the relativistic (Lorentz) invariance fixes the spectral density eq.$\left( 
\ref{spectrum}\right) $ modulo the scale factor $
\rlap{\protect\rule[1.1ex]{.325em}{.1ex}}h%
$ that should be identified with Planck constant in order to agree with the
predictions of QM.

\textbf{Excited pure states} of the field correspond in HS to vectors which
may be obtained by application of the creation operators of photons $\left\{ 
\hat{a}_{j}^{\dagger }\right\} $ on the vacuum state. For instance $\hat{a}%
_{j}\dagger \mid 0\rangle $ is the vector of HS representing the state of a
single photon of kind $j$, the vector $\hat{a}_{j}^{\dagger }\hat{a}%
_{k}^{\dagger }\mid 0\rangle $ represents a two-photon state, etc. The set
of all these states is named Fock space. It is usual to admit that any
linear combinations of state-vectors belonging to Fock space is also a
possible pure state. Hence any vector in HS got by the action of a function
of the creation operators on the vacuum state, leads to a pure state $\mid
\psi \rangle ,$ that is 
\begin{equation}
\mid \psi \rangle =\hat{f}\mid 0\rangle ,\hat{f}\equiv
c_{0}+\sum_{j}\sum_{n}c_{jn}\hat{a}_{j}^{\dagger n},\langle \psi \mid \psi
\rangle =1,  \label{stateHS}
\end{equation}
where $j$ labels a radiation mode and $n$ is a natural number, $c_{0}$ and $%
c_{jn}$ being complex numbers, with the constraint that the vector $\mid
\psi \rangle $ is normalized. Mixed states, which might be represented by
density operators, correspond to probability distributions of pure states.

For all states defined by eq.$\left( \ref{stateHS}\right) $ in the HS
formalism we may got the corresponding expression in WW via performing the
(inverse) Weyl transform, similarly as we made for the vacuum state going
from eq.$\left( \ref{2d}\right) $ to eq.$\left( \ref{1}\right) .$ That is
the set of states in WW will consist of functions of the amplitudes with the
form 
\begin{equation}
\Psi \left( \left\{ a_{j},a_{j}^{*}\right\} ,t\right) =T_{W}\left[ \hat{f}%
\mid 0\rangle \langle 0\mid \hat{f}^{\dagger }\right] .  \label{stateWW}
\end{equation}
Working these transforms is straightforward although lengthy in most cases.
For the Fock states $\left\{ \hat{a}_{j}^{\dagger n}\mid 0\rangle \right\} $
the results are well known but will not be reproduced here.

I shall point out that the definition of states in HS is controversial. In
fact the \textit{question whether a given state of the set eq.}$\left( \ref
{stateHS}\right) $\textit{\ is physical} cannot be answered within the nude
HS formalism. It would require the existence of a ``preparation procedure''
able, in principle, to manufacture the said state in the laboratory. For
some states belonging to the set eq\textit{.}$\left( \ref{stateHS}\right) $
that preparation seems not possible because some of the state properties are
contradictory. For instance the state $\hat{a}_{j}^{\dagger }\mid 0\rangle $
is not localized but \textit{extended }over the whole space (or over a large
normalization volume) if $\hat{a}_{j}^{\dagger }$ is associated to a plane
wave with definite wavevector $\mathbf{k}_{j}$. But at the same time it
should represent a (\textit{localized?)} particle, i.e. one ``photon''. Thus
many authors assume that only a fraction of the set eq\textit{.}$\left( \ref
{stateHS}\right) $ are physical states, e.g. those in the form of
(localized) wavepackets.

\textbf{Observables. }Eq.$\left( \ref{Wtransform}\right) $ allows getting
the WW counterparts of the observables in the HS formalism. They would be
functions of the amplitudes $\left\{ a_{j},a_{j}^{*}\right\} .$

\textbf{Evolution. }It is remarkable that the evolution\textbf{\ }of the 
\textit{quantized} \textit{free EM field} within the WW formalism is just 
\textit{the classical (Maxwell) evolution}. In fact the evolution of a field
in the WW formalism is given by the Moyal equation \cite{Scully}, \cite
{Zachos}, \cite{EPJP} 
\begin{eqnarray}
\frac{\partial W_{\hat{M}}}{\partial t} &=&2\{\sin \left[ \frac{1}{4}\left( 
\frac{\partial }{\partial \mathrm{Re}a_{j}^{\prime }}\frac{\partial }{%
\partial \mathrm{Im}a_{j}^{\prime \prime }}-\frac{\partial }{\partial 
\mathrm{Im}a_{j}^{\prime }}\frac{\partial }{\partial \mathrm{Re}%
a_{j}^{\prime \prime }}\right) \right] \smallskip  \nonumber \\
&&\times W_{\hat{M}}\left\{ a_{j}^{\prime },a_{j}^{*\prime },t\right\}
H\left( a_{j}^{\prime \prime },a_{j}^{*\prime \prime }\right)
\}_{a_{j}},\smallskip  \label{Moyal}
\end{eqnarray}
where $\left\{ {}\right\} _{a_{j}}$ means making $a_{j}^{\prime
}=a_{j}^{\prime \prime }=a_{j}$ and $a_{j}^{*\prime }=a_{j}^{*\prime \prime
}=a_{j}^{*}$ after performing the derivatives. The density operator $\hat{M}$
is the representative of a state in HS and $W_{\hat{M}}$ its Wigner
function. For the free electromagnetic field the Hamiltonian $H\left(
a_{j},a_{j}^{*}\right) $ is quadratic in the amplitudes (see eq.$\left( \ref
{Hamiltonians}\right) $ below$)$ whence $\sin x$ reduces to $x$ in eq.$%
\left( \ref{Moyal}\right) $. This means that the Moyal equation becomes 
\begin{eqnarray}
\frac{\partial W_{\hat{M}}}{\partial t} &=&\frac{1}{2}\left( \frac{\partial 
}{\partial \mathrm{Re}a_{j}^{\prime }}\frac{\partial }{\partial \mathrm{Im}%
a_{j}^{\prime \prime }}-\frac{\partial }{\partial \mathrm{Im}a_{j}^{\prime }}%
\frac{\partial }{\partial \mathrm{Re}a_{j}^{\prime \prime }}\right)
\smallskip  \nonumber \\
&&\times W_{\hat{M}}\left\{ a_{j}^{\prime },a_{j}^{*\prime },t\right\}
H\left( a_{j}^{\prime \prime },a_{j}^{*\prime \prime }\right)
\}_{a_{j}},\smallskip  \label{Poisson}
\end{eqnarray}
where the right side is a Poisson bracket, taking eqs.$\left( \ref{10}%
\right) $ into account, and eq.$\left( \ref{Poisson}\right) $ is the
Liouville equation of classical evolution. For details see e.g. Ref. \cite
{EPJP} .

\textbf{Interactions. }Studying, within WW, the evolution when the EM field
is not free requires the inclusion of the interactions with other fields or
with macroscopic bodies. The interaction with other fields should be made
also within a WW representation for the other fields. See comments on this
possibility at the end of section 3 below. In particular we cannot study QED
within the WW approach, because this would require the Dirac
electron-positron field in a formalism compatible with WW of the EM field.
This difficulty reduces dramatically the possibilities of our approach.

In sharp contrast it is easy to define the interactions of the EM field with
macroscopic bodies, which allows the study of a interesting experiments. A
relevant example is worked in section 4. I propose that the interactions of
the quantized EM field, in the WW representation, with macroscopic bodies
are given by classical Maxwell-Lorentz electrodynamics. The assumption is
plausible because both macroscopic bodies and the free EM field are governed
by classical laws, that is eq.$\left( \ref{Poisson}\right) $.

\textbf{Hamiltonian. }The free field Hamiltonians in the HS and WW
formalisms might be defined as follows 
\begin{equation}
\hat{H}_{HS}=
\rlap{\protect\rule[1.1ex]{.325em}{.1ex}}h%
\sum_{j}\omega _{j}(\hat{a}_{j}^{\dagger }\hat{a}_{j}+\frac{1}{2})=\frac{1}{2%
}
\rlap{\protect\rule[1.1ex]{.325em}{.1ex}}h%
\sum_{j}\omega _{j}(\hat{a}_{j}^{\dagger }\hat{a}_{j}+\hat{a}_{j}\hat{a}%
_{j}^{\dagger }),H_{WW}=
\rlap{\protect\rule[1.1ex]{.325em}{.1ex}}h%
\sum_{j}\omega _{j}\left| a_{j}\right| ^{2},  \label{Hamiltonians}
\end{equation}
respectively, where I have taken eq.$\left( \ref{sym}\right) $ into account.
However with this definition the energy of the vacuum state $\mid 0\rangle $
is not zero (in fact the vacuum becomes the ZPF with distribution eq.$\left( 
\ref{1}\right) $) whence it is standard practice to redefine the Hamiltonian
by putting the annihilation operators on the right, which is named `normal
ordering rule'. Thus the HS and WW Hamiltonians become, respectivly, 
\begin{equation}
\hat{H}_{HS}^{norm}=
\rlap{\protect\rule[1.1ex]{.325em}{.1ex}}h%
\sum_{j}\omega _{j}\hat{a}_{j}^{\dagger }\hat{a}_{j},H_{WW}^{norm}= 
\rlap{\protect\rule[1.1ex]{.325em}{.1ex}}h%
\sum_{j}\omega _{j}\left( \left| a_{j}\right| ^{2}-\frac{1}{2}\right)
.\smallskip  \label{rule}
\end{equation}
Actually the normal order produces the same effect on the field energy than
subtracting from the Hamiltonian eq.$\left( \ref{Hamiltonians}\right) $ its
vacuum expectation value , that is 
\begin{equation}
\hat{H}_{HS}^{subtract}=\hat{H}_{HS}-\langle 0\mid \hat{H}_{HS}\mid 0\rangle
=
\rlap{\protect\rule[1.1ex]{.325em}{.1ex}}h%
\sum_{j}\omega _{j}\hat{a}_{j}^{\dagger }\hat{a}_{j},  \label{subtract}
\end{equation}
also giving nil vacuum energy.

When there are interactions with other quantized fields the Hamiltonian
cannot be defined prior to obtaining formalisms similar to WW for those
fields, which at the moment unavailable. For the case when the quantized EM
field interacts with macroscopic bodies I propose that the interaction
Hamiltonian is given by classical electrodynamics.

\textbf{Vacuum energy expectation value. }Let us calculate the expectation
value of the energy within WW for a general state $W_{\phi }$ . In order to
agree with HS predictions we shall use the normally ordered Hamiltonian eq.$%
\left( \ref{Hnormal}\right) .$ For a state $\mid \phi \rangle $ of the field
the calculation, firstly in HS then in WW, may be written (see eq.$\left( 
\ref{1a}\right) )$%
\begin{eqnarray}
\left\langle E\right\rangle &=&\left\langle \phi \left| \sum_{l} 
\rlap{\protect\rule[1.1ex]{.325em}{.1ex}}h%
\omega _{l}\hat{a}_{l}^{\dagger }\hat{a}_{l}\right| \phi \right\rangle 
\nonumber \\
&=&\sum_{l}
\rlap{\protect\rule[1.1ex]{.325em}{.1ex}}h%
\omega _{l}\int W_{\phi }\left( \left\{ a_{l}\right\} \right) \left( \left|
a_{l}\right| ^{2}-\frac{1}{2}\right) d\func{Re}a_{l}d\func{Im}a_{l} 
\nonumber \\
&=&\sum_{l}
\rlap{\protect\rule[1.1ex]{.325em}{.1ex}}h%
\omega _{l}\int \left| a_{l}\right| ^{2}\left( W_{\phi }\left( \left\{
a_{l}\right\} \right) -W_{0}\left( \left\{ a_{l}\right\} \right) \right) d%
\func{Re}a_{l}d\func{Im}a_{l},\smallskip  \label{energy}
\end{eqnarray}
where $W_{\phi }\left( \left\{ a_{l}\right\} \right) $ is the state (Wigner
function) that in WW represents the HS state $\mid \phi \rangle $ and $W_{0}$
is the vacuum Wigner function eq.$\left( \ref{1}\right) .$ The result is
that the \textit{normal ordering rule of HS corresponds to removing the
vacuum energy} that would appear according to the initial definition eq.$%
\left( \ref{Hamiltonians}\right) $. Thus the vacuum Wigner function eq.$%
\left( \ref{1}\right) $ plays a strange role in the WW representation. On
the one hand the vacuum eq.$\left( \ref{1}\right) $ does not imply the
absence of any field, whence the vacuum field should interact with charges.
On the other hand the vacuum is devoid of energy according to eq.$\left( \ref
{energy}\right) ,$ then appearing as an inert stuff without any physical
relevance, except for its fluctuations. This paradoxical condition will be
studied in more detail in later sections.

\textbf{Expectation value }of the observable $\hat{O}$ in the state $\hat{%
\rho}$ reads $Tr(\hat{\rho}\hat{O})$, or in particular $\langle \psi \mid 
\hat{O}\mid \psi \rangle ,$ in the HS formalism. The counterpart in the WW
formalism leads to the integral of the product of two functions of the
amplitudes, that is 
\begin{equation}
Tr(\hat{\rho}\hat{O})=\int W_{\hat{\rho}}\left\{ a_{j},a_{j}^{*}\right\} W_{%
\hat{O}}\left\{ a_{j},a_{j}^{*}\right\} \prod_{j}d\mathrm{Re}a_{j}d\mathrm{Im%
}a_{j},  \label{expect}
\end{equation}
where $W_{\hat{\rho}}$ and $W_{\hat{O}}$ are the counterparts of a density
operator, $\hat{\rho},$ and a quantum observable $\hat{O}.$ A particular
case of eq.$\left( \ref{expect}\right) $ is the vacuum expectation value
when $W_{\hat{\rho}}$ becomes $W_{0}.$

In summary eq.$\left( \ref{expect}\right) $ proves that \textit{the
predictions of both formalisms, HS and WW, are identical for the same states
and observables. This implies that both correspond to the same physical
theory for the quantized EM field, although the mathematical formalisms are
quite different.}

\section{The WW formalism as a new field quantization method}

In the previous section I have shown that WW is a new formalism for the
quantized EM field, equivalent to the (HS) standard one. The equivalence
means that we may represent states and observables in either formalism,
passing from WW to HS via the Weyl transform (not written explicitly in this
paper, see e.g. \cite{FOOP}) or from HS to WW via the inverse transform eq.$%
\left( \ref{Wtransform}\right) .$ The point is that expectation values may
be got in either HS or WW giving the same numerical values in both cases.

In this section I propose to take WW as a new quantization method
alternative to the (HS) standard one. That is I will skip the passage
through HS in order to get WW, as made in the previous section. In fact I
will quantize directly the classical (Maxwell) EM theory via a set of rules
shown below. I will use that method for the EM field interacting with
macroscopic bodies, these characterized by their charges or their bulk
electric and magnetic properties. This includes most of the devices used in
quantum optics like mirrors, polarizers, beam splitters, etc. The motion of
the macroscopic bodies is assumed non-relativistic.

The new quantization method of the EM field may be stated by the following
rules:

1. The formalism rests on spacetime functions as in the classical Maxwell
theory. In particular the field is defined in the Coulomb gauge by a vector
potential \textbf{A(r},t), eq.$\left( \ref{A}\right) ,$ and a scalar
potential $\phi \left( \mathbf{r},t\right) $ whence we might derive the
electric \textbf{E(r},t) and the magnetic \textbf{B(r},t) fields. The vector
potential may be expanded in normal modes and written in terms of amplitudes 
$\left\{ a_{j},a_{j}^{*}\right\} $ if necessary.

2. The evolution of the field is governed by classical Maxwell-Lorentz
electrodynamics and the motion of macroscopic bodies by the standard laws of
mechanics, that is the Newton's second law or Hamiltonian dynamics.

3. The \textit{essential difference} between classical electrodynamics and
the quantum theory of the EM field in WW \textit{is\ just} \textit{the
assumption that there is a random EM field filling space, }i.e. the ZPF with
the probability distribution eq.$\left( \ref{1}\right) .$ Actually that
distribution might be obtained without any reference to the HS formalism. In
fact eqs.$\left( \ref{1}\right) $ and $\left( \ref{spectrum}\right) $ may be
derived from the following two conditions put on the ZPF distribution,
except for a scaling parameter. That is maximal information (Shanon) entropy
compatible with relativistic (Lorentz) invariance of the spectrum. The
scaling parameter that appears should be identified with Planck constant 
\rlap{\protect\rule[1.1ex]{.325em}{.1ex}}h%
.

4. The Hamiltonian of the free field is given as a function of the
amplitudes by 
\begin{equation}
H=
\rlap{\protect\rule[1.1ex]{.325em}{.1ex}}h%
\sum_{j}\omega _{j}\left| a_{j}\right| ^{2}-\left\langle H\right\rangle
_{vac}=
\rlap{\protect\rule[1.1ex]{.325em}{.1ex}}h%
\sum_{j}\omega _{j}\left| a_{j}\right| ^{2}-\frac{1}{2} 
\rlap{\protect\rule[1.1ex]{.325em}{.1ex}}h%
\sum_{j}\omega _{j},  \label{H}
\end{equation}
where $\left\langle H\right\rangle _{vac}$ is the part coming from the ZPF
which must be subtracted, see eq.$\left( \ref{1a}\right) ,$ so that the
energy of the ground state is zero. The Hamiltonian including interactions
with macroscopic bodies should be defined by eq.$\left( \ref{H}\right) $
plus the appropriate interaction terms, according to classical
electrodynamics.

I point out that the assumption of a random radiation filling space (the
ZPF) in \textit{rule 3} and eq.$\left( \ref{H}\right) $ of \textit{rule 4}
look inconsistent. Indeed the ZPF is assumed to act on electric charges, but
itself is devoid of energy. Of course eq.$\left( \ref{H}\right) $ is related
to the ``normal ordering rule'', which in standard (HS) quantum theory looks
artificial but not inconsistent. The label of inconsistence may be overthrow
by taking into account that our approach has a limited domain, namely the EM
field plus macroscopic bodies. We may assume that in a wider domain that
included all fields (i.e. those of the standard model of high energy
physics) and interactions, the total energy of the ZPF would be plausibly
nil. Indeed we may assume that (negative) contributions of Fermi fields
should balance the (positive) contributions of Bose fields. In the meantime,
when a theory similar to WW does not exists for all fields, it is
appropriate to remove the ZPF contribution of the EM field in calculations.

5. The set of states are functions $\Psi \left( \left\{
a_{j},a_{j}^{*}\right\} \right) $ of the amplitudes (or functions of the
electric \textbf{E} and magnetic \textbf{B }fields). However I will not fix
the set of states as given in eq.$\left( \ref{stateWW}\right) $ of section
2, but I will define the possible states as those probability distributions
(given by positive definite functions) of the EM field. I point out that
this definition of states might break the full agreement between WW and HS
derived in section 2. However I believe that both formulations could not be
discriminated empirically.

6. In principle all functions of the amplitudes, F$\left( \left\{
a_{j},a_{j}^{*}\right\} \right) $ ( or functions of the electric \textbf{E}
and magnetic \textbf{B }fields) may be observables.

7. The expectation value of the observable $F\left( \left\{
a_{j},a_{j}^{*}\right\} \right) $ in the state $\Psi \left( \left\{
a_{j},a_{j}^{*}\right\} \right) $ may be obtained via the integral 
\begin{equation}
\int F\left\{ a_{j},a_{j}^{*}\right\} \Psi \left\{ a_{j},a_{j}^{*}\right\}
\prod_{j}d\mathrm{Re}a_{j}d\mathrm{Im}a_{j}.  \label{expec}
\end{equation}

I shall point out that in sections 2 and 3 I have ignored all effects of the
scalar potential $\phi $(\textbf{r},t), that appears when we work in the
Coulomb gauge, deriving from the instantaneous electrostatic interactions.
We must also take into account that potential in the interpretation of
experiments, which would be straightforward.

A most relevant question is whether a quantization formalism similar to the
WW one studied here may be applied to other quantum fields. My answer is
that I do not know, but certainly the search for this possibility is
worthwhile. I believe that the generalization of WW to Bose fields is likely
possible because the essential properties needed are fulfilled. In fact the
fields are represented by ordinary scalar or vector functions of the
coordinates, they may be expanded in normal modes (in particular plane
waves) and the amplitudes may be quantized via promoting the amplitudes of
the modes to be operators fulfilling the standard commutation rules eqs.$%
\left( \ref{com}\right) .$ Of course I exclude the gravitational field,
which being nonlinear cannot be quantized that way. In sharp contrast the
standard quantization of Fermi fields involves field amplitudes fulfilling
anticommutation rules (e.g. the Dirac electron-positron field). Furthermore
in that case the amplitudes are not simple numerical (scalar or vector)
functions of the coordinates, but spinors with several components. Therefore
finding a quantization method similar to WW will be substantially more
difficult if at all possible.

\section{Entangled photons experiments}

In this section I will study a relevant experiment within the WW
quantization formalism, that is an optical test of the Bell inequality \cite
{Bell} involving parametric down conversion. The experiment embraces the EM
field interacting with macroscopic bodies. Therefore the interactions will
be treated as in the classical Maxwell-Lorentz electrodynamics, as said in
the previous section. I will study an experiment similar to the one reported
in Refs. \cite{Shalm},\cite{Giustina}, \cite{BIG} but in order to simplify
the argument I shall consider maximally entangled photons, although
non-maximally entangled photons were used in the quoted test.

In experiments with ``entangled photon pairs'' produced via spontaneous
parametric down-conversion (SPDC) a crystal having nonlinear electric
susceptibility is pumped by a laser with frequency $\omega _{P}$. In the
opposite side of the crystal a rainbow with several colors appears. Amongst
the radiation emitted, two narrow beams, named ``signal'' and ``idler'', are
selected via apertures and lens systems. For two appropriate (conjugated)
directions these beams are strongly correlated, which in the HS formalism is
labeled entanglement. Indeed in HS the two beams are interpreted as
consisting of a flow of entangled photon pairs, one photon in every pair
going to the ``signal'' beam and the partner photon to the ``idler'' beam.
The energy of one photon from the laser is $
\rlap{\protect\rule[1.1ex]{.325em}{.1ex}}h%
\omega _{P}$ and the SPDC process is currently interpreted saying that one
of the laser photons is split by the coupling with the crystal, giving rise
to two photons with energies $
\rlap{\protect\rule[1.1ex]{.325em}{.1ex}}h%
\omega _{s}$ and $
\rlap{\protect\rule[1.1ex]{.325em}{.1ex}}h%
\omega _{i},$ fulfilling $\omega _{s}$+$\omega _{i}=\omega _{P}$, this
equality interpreted as energy conservation. In contrast the photon momenta
are not conserved because a part of the momentum is absorbed by the crystal.
This is the common view but I stress that in the WW representation there are
no photons, just (continuous) wave fields.

SPDC is actually a process that may produced at the macroscopic level, when
it can be interpreted classically \cite{Kaled}. In the process two
macroscopic light signals with wavevectors $\mathbf{k}_{s}$ and $\mathbf{k}%
_{i},$ and frequencies $\omega _{s}$ and $\omega _{i}$ respectively, plus a
strong laser beam with frequency $\omega _{P}=\omega _{s}$+$\omega _{i}$ are
sent to a nonlinear crystal. For appropriate $\mathbf{k}_{s}$ and $\mathbf{k}%
_{i}$ two beams appear on the opposite side of the crystal with the same
wavevectors and frequencies as the incoming ones, but greater intensities,
which may be represented as follows 
\begin{eqnarray}
incoming &:&a_{s}\exp \left( i\mathbf{k}_{s}\mathbf{\cdot r-}i\omega
_{s}t\right) ,a_{i}\exp \left( i\mathbf{k}_{i}\mathbf{\cdot r-}i\omega
_{i}t\right) ,  \nonumber \\
outgoing &:&[a_{s}+Da_{i}^{*}\exp (i\zeta )]\exp \left( i\mathbf{k}_{s}%
\mathbf{\cdot r-}i\omega _{s}t\right) ,  \nonumber \\
&&[a_{i}+Da_{s}^{*}\exp \left( i\xi \right) ]\exp \left( i\mathbf{k}_{i}%
\mathbf{\cdot r-}i\omega _{i}t\right) ,  \label{3.0}
\end{eqnarray}
where $D$ is a small coupling parameter, i. e. $0<D$ $<<1$, $\zeta $ and $%
\xi $ are phases of the outgoing beams, those in the incoming beams being
included in the definition of the amplitudes $a_{s}$ and $a_{i}$. The
remaining notation is standard. An actual light beam consists of a
superposition of many radiation modes with close wavevectors. However I will
use a ``two-modes approximation'', which justifies using sharp wavevectors
in eqs.$\left( \ref{3.0}\right) .$

Within classical electrodynamics the SPDC process may be described as
follows \cite{Kaled}. The superposition of the incoming beam with frequency $%
\omega _{s}$, that I will label ``signal'', and the laser with frequency $%
\omega _{P}$ produces a resonant vibration of some elements (electrons) of
the crystal, which radiate with frequency $\omega _{P}-\omega _{s}=\omega
_{i}$ mainly in a particular direction, which should agree with the
direction that we chose for the other incoming beam, labeled ``idler''.
Similarly the incoming idler beam with frequency $\omega _{i}$ toghether
with the laser gives rise to radiation with frequency $\omega _{s}$ in the
direction of the incoming signal beam. Thus the equality $\omega _{P}=\omega
_{s}$+$\omega _{i}$ is just a matching condition between the laser, the
incoming fields and some vibration modes of the crystal. The relevant point
is that the amplitude $a_{i}^{*}$ that is superposed to the amplitude $a_{s}$
of the ``signal'' is precisely the complex conjugate of the amplitude of the
incoming ``idler'' beam, and similarly for the other superposition, see eqs.$%
\left( \ref{3.0}\right) $. This is the reason for the strong correlation
between the outgoing ``signal'' and ``idler'' beams.

Now we may study within the WW the SPDC phenomenon involved in tests of Bell
inequalities. It is similar to the macroscopic SPDC above described, but in
this case the incoming beams belong to ZPF radiation eq.$\left( \ref{1}%
\right) $. In practice SPDC is used in the optical range, i.e. the EM fields
involved belong to the visible or near visible light. As said above the
strong correlation between the two outgoing beams is named ``entanglement''
in the standard HS quantum formalism. Indeed in the HS formalism the beams
are viewed as two flows of pairs of entangled photons, as said above. It is
common to stress that entanglement is a typical quantum phenomenon, quite
different from any classical correlation. In WW we see that it is
``quantum'' because it involves the random radiation ZPF, eq.$\left( \ref{1}%
\right) ,$ which is indeed the characteristic trait of the EM quantized
field as shown in section 3.

In practice the effects of the spacetime dependence and the extra phases of
eqs.$\left( \ref{3.0}\right) $ play no role in most cases, and they can be
ignored, writing simply 
\begin{equation}
incoming:a_{s},a_{i},outgoing:(a_{s}+Da_{i}^{*}),(a_{i}+Da_{s}^{*}).
\label{31}
\end{equation}
However sometimes the phases are relevant and they shall be taken into
account. With suitable devices the signal and idler beams are combined
giving rise to two beams strongly correlated in polarization. These beams
travel (maybe a long path) until Alice and Bob respectively. Alice possesses
a polarization analyzer and a detector in front of it, and similarly Bob.
Thus the beams arriving at these two detectors may be represented, in the
two modes approximation, by the field amplitudes \ref{book} 
\begin{eqnarray}
A &=&a_{s}\cos \theta +ia_{i}\sin \theta +D[a_{i}^{*}\cos \theta
+ia_{s}^{*}\sin \theta ]\equiv A_{0}+DA_{1},  \nonumber \\
B &=&a_{i}\cos \phi -ia_{s}\sin \phi +D[a_{s}^{*}\cos \phi -ia_{i}^{*}\sin
\phi ]\equiv B_{0}+DB_{1},  \label{3.2}
\end{eqnarray}
where $\theta $ and $\phi $ are the polarizer\'{}s angles, A and B standing
from Alice and Bob respectively. The amplitudes complex conjugate to $A$ and 
$B$ will be labelled $A^{*}$and $B^{*}$ respectively. From eqs.$\left( \ref
{3.2}\right) $ it is straightforward to get the quantum predictions for the
experiment within WW. We should take into account that 
\begin{equation}
\langle \left| a_{s}\right| ^{2}\rangle =\langle \left| a_{i}\right|
^{2}\rangle =\langle \left| a_{s}\right| ^{4}\rangle =\langle \left|
a_{i}\right| ^{4}\rangle =\frac{1}{2},\left\langle
a_{s}^{m}a_{s}^{*n}\right\rangle =0\text{ if }m\neq n,  \label{6.3}
\end{equation}
where the averages, noted $\left\langle {}\right\rangle ,$ correspond to
integrals involving W$_{0,}$ eq.$\left( \ref{1}\right) $.

A rigorous theory of detection would be complex and it will not be attempted
here. I shall just make an extremely simple proposal, at the same time
plausible and fitting with the common approach in the standard HS
treatments. That is I will assume that both Alice and Bob single detection
rates are proportional to the radiation intensity arriving at their
detectors respectively, with the ZPF (vacuum) intensity subtracted. That
intensity is just what would arrive at the detectors if the pumping laser
was switched off, and the subtraction amounts to putting D=0 in eq.$\left( 
\ref{3.0}\right) .$ Modulo a proportionality constant (assumed equal for
Alice and Bob) we get for the Alice detection rate 
\begin{equation}
P_{A}=\langle \mid A\mid ^{2}\rangle -\langle \mid A_{0}\mid ^{2}\rangle
=D^{2}\langle \mid A_{1}\mid ^{2}\rangle +2D\langle \func{Re}\left(
A_{0}A_{1}^{*}\right) \rangle .  \label{3.3}
\end{equation}
In the following I will interpret $P_{A}$ eq.$\left( \ref{3.3}\right) $ as
the probability of a detection event within a given time window short enough
so that more than one detection event is unkikely within the window. Thus
the detection rate will be the product of $P_{A}$ times the rate of time
windows.

The second term of eq.$\left( \ref{3.3}\right) $ does not contribute because
it is a linear combination of the averages $\left\langle a^{2}\right\rangle $
= $\left\langle a^{*2}\right\rangle =0$ (see eq.$\left( \ref{6.3}\right) )$
whence we get 
\begin{eqnarray}
P_{A} &=&D^{2}\langle \mid A_{1}\mid ^{2}\rangle =D^{2}\langle \left|
(a_{i}\cos \theta -ia_{s}\sin \theta )(a_{i}^{*}\cos \theta +ia_{s}^{*}\sin
\theta )\right| \rangle  \label{3.4} \\
&=&D^{2}\left( \left\langle \left| a_{i}\right| \right\rangle \cos
^{2}\theta +\left\langle \left| a_{s}\right| \right\rangle \sin ^{2}\theta
\right) =\frac{1}{2}\left| D\right| ^{2},  \nonumber
\end{eqnarray}
A similar result may be obtained for the single detection rate of Bob, that
is 
\begin{equation}
P_{A}=P_{B}=\frac{1}{2}\left| D\right| ^{2}.  \label{3.5}
\end{equation}

The coincidence detection rate should obviously depend on the correlation
between the signals arriving at Alice and Bob respectively. However it is
not trivial to get a quantitative proposal from this qualitative idea,
because some ambiguity arises. We might assume that the coincidence
detection probability depends on the intensity correlation with the ZPF
subtracted, but this would ignore the phase correlations, that in this case
are relevant. Thus I propose that the coincidence probability in a given
detection window, $P_{AB}$, will depend on the correlation of the amplitudes
A and B, eqs.$\left( \ref{3.2}\right) ,$ arriving a Alice and Bob
respectively, as follows 
\begin{equation}
P_{AB}=\left| \left\langle AB\right\rangle \right| ^{2},  \label{5.0}
\end{equation}
which is consistent with the single detection eq.$\left( \ref{3.3}\right) .$
Here the vacuum subtraction is not needed because the detection probability
with the pumping laser switched off gives no contribution. In fact, from eqs.%
$\left( \ref{6.3}\right) $ we get, 
\[
\left\langle A_{0}B_{0}\right\rangle =0\Rightarrow \left| \left\langle
A_{0}B_{0}\right\rangle \right| ^{2}=0. 
\]
Taking eqs.$\left( \ref{3.2}\right) $ into account we obtain 
\begin{eqnarray}
\left\langle AB\right\rangle &=&\langle \left( A_{0}+DA_{1}\right) \left(
B_{0}+DB_{1}\right) \rangle  \nonumber \\
&=&\left\langle A_{0}B_{0}\rangle +D\langle A_{0}B_{1}\rangle +D\langle
A_{1}B_{0}\rangle +D^{2}\langle A_{1}B_{1}\right\rangle .  \label{5.1}
\end{eqnarray}
The first and last terms are nil according to eqs.$\left( \ref{6.3}\right) $%
. Thus we tentatively obtain 
\begin{equation}
P_{AB}=\left| \langle AB\rangle \right| ^{2}=D^{2}\left| \langle
A_{0}B_{1}\rangle +\langle A_{1}B_{0}\rangle \exp \left( i\chi \right)
\right| ^{2},  \label{5.2}
\end{equation}
where in eq.$\left( \ref{5.2}\right) $ I have restored the phases which were
ignored in eq.$\left( \ref{5.1}\right) $ (see comment below eq.$\left( \ref
{3.0}\right) ).$ Indeed it is plausible that the phases of the two terms of
eq.$\left( \ref{5.2}\right) $ are different and both may be assumed random
variables with a homogeneous distribution in $\left[ 0,2\pi \right] $ each.
Thus averaging over the relative phase $\chi $ in eq.$\left( \ref{5.2}%
\right) $ we get 
\[
P_{AB}=D^{2}(\left| \langle A_{0}B_{1}\rangle \right| ^{2}+\left| \langle
A_{0}B_{1}\rangle \right| ^{2}). 
\]
These averages may be got taking eqs.$\left( \ref{3.2}\right) $ into account
and we obtain 
\begin{eqnarray*}
\langle A_{0}B_{1}\rangle &=&\left\langle \left| a_{s}\right| ^{2}\cos
\theta \cos \phi +\left| a_{i}\right| ^{2}\sin \theta \sin \phi
\right\rangle =\frac{1}{2}\cos \left( \theta -\phi \right) , \\
\langle A_{1}B_{0}\rangle &=&\left\langle \left| a_{i}\right| ^{2}\cos
\theta \cos \phi +\left| a_{s}\right| ^{2}\sin \theta \sin \phi
\right\rangle =\frac{1}{2}\cos \left( \theta -\phi \right) .
\end{eqnarray*}
whence the coincidence probability will be

\begin{equation}
P_{AB}=\frac{1}{2}D^{2}\cos ^{2}\left( \theta -\phi \right) .  \label{5.4}
\end{equation}
The results eqs.$\left( \ref{3.5}\right) $ and $\left( \ref{5.4}\right) $
violate a Bell inequality \cite{CH}. The result is arrived at via classical
EM and ZPF, and then spooky action at a distance is not required to violate
the Bell inequalities in this model.

\section{Discussion}

I have shown, in section 3 of this article, that the classical (Maxwell)
electromagnetic field may be quantized just assuming the existence of a
random radiation filling space, the ``zeropoint field'' (ZPF). The ZPF is
quantitatively characterized by the distribution eq.$\left( \ref{W0}\right) $
of energies amongst the normal modes of the field$.$ The formalism obtained
by this method may be labeled ``Weyl-Wigner representation'' (WW) of the
quantum EM field. For a complete specification and understanding of WW we
must recall the following answers to three relevant questions.

The first question is whether WW is equivalent to the standard Hilbert space
(HS) quantum theory of the EM field. The answer is affirmative provided that
we assume that the ZPF contributes nil energy to the free field (i.e. the
field without interactions). Thus the ZPF represents the ``vacuum state''.
The proof of the assertion is that there is a reversible Weyl transform that
allows passing from HS to WW and from WW to HS, as studied in section 2.

The second question is whether WW may be interpreted as local realistic,
like classical theories. The answer might be also affirmative provided the
following two points are clarified. Firstly we must explain the paradoxical
result that the strong ZPF possesses nil energy, what I have made in section
3. Secondly we must make a change in the formalism of section 2, assuming
now that the set of states in WW corresponds to \textit{positive }functions
of the amplitudes, so that these functions may be interpreted as probability
distributions . In section 3 I have argued that this change from the states
in HS (mixtures of Fock states) to WW states, likely does not disturb the
agreement of WW with experiments. Therefore even if HS and WW are not fully
equivalent both allow a fair interpretation of the experiments, and WW has
the advantage that it permits a realistic local understanding.

The third question is whether the WW quantization of the EM field may be
extended to other fields, e.g. the Dirac electron-positron field. My answer
is that I do not know, but I believe that the answer is affirmative for Bose
fields. Certainly the absence of a generalization to other fields is a big
difficulty that requires further research work. But at least the field in WW
may be coupled to macroscopic bodies maintaining the possibility of a local
realistic interpretation. This allows the interpretation of relevant
experiments as shown in section 4. Namely the crucial (``loophole free'')
tests of Bell inequalities \cite{Shalm}, \cite{Giustina}, \cite{BIG} which
proves that they may be interpreted maintaining locality. Those experiments
had been taken as the death of local realism \cite{Wiseman}, \cite{Aspect}, 
\cite{Oxford}. Further discussion about the results of section 4 will be
provided below.

The WW formalism for the EM field is a translation of the old Wigner
representation for quantum mechanics of particles (QM) in terms of functions
in phase space. I believe that such translation has not been worked in
detail for fields due to the big obstacle that the Wigner representation in
QM does not admit a realistic interpretation because the Wigner functions
are not positive in general. I argue that the case for the EM field is quite
different. Indeed, as said above the formalism of section 3 admits a local
realistic interpretation in terms of fields without the need of particles
(photons), which have a rather counterintuitive behaviour as commented in
section 2.

An interesting question is why the Wigner representation of QM (for systems
of particles) does not admits a realistic interpretation. I believe that the
reason is that quantum particles are dramatically different from classical
particles. The latter may be treated as lacking structure, so that positions 
$\left\{ \mathbf{r}_{j}\right\} $ and momenta $\left\{ \mathbf{p}%
_{j}\right\} $ determine the state of a system of particles at a given time.
They determine both the past of the future given the interactions, which
exemplifies the celebrated Laplacian determinism of classical mechanics.
Hence the path of the particles is a line in phase space parametrized by
time, which explains the relevance of phase space in classical mechanics. In
contrast quantum particles are complex structures dressed with many fields,
in particular the ZPF's. Hence the particles motion is governed by both the
external forces and the dressing fields. Thus an interesting research is to
attempt deriving the laws of motion of the particles from the properties of
the fields plus the given forces. In the case of charged particles it is
plausible that the main contribution comes from the EM field, in particular
the ZPF. Indeed it is the case that a line of research on the subject has
been alive from long ago, with the name of \textit{stochastic electroynamics}
\cite{Braffort}, \cite{Marshall}, \cite{Dice}, \cite{Dice2}, \cite{Boyer}, 
\cite{book}. This is a theory that attempts explaining phenomena taken as
purely quantal, from the action on charged particles of an assumed EM random
radiation field. Within this line a recent article presents a derivation of
the canonical commutation rules \cite{Cetto}.

As a final part of this section I shall discuss whether the analysis of
section 4 does provide a local realistic interpretation of the tests of Bell
inequalities. Most of the development is a straightforward application of
the WW formalism that fits in local realism as exposed in section 3, but the
photodetection involves extra assumptions that require discussion. Let us
start with the single detection. The assumption that the probability of
detection, in one time window, is proportional to the arriving intensity is
plausible and follows the usual practice. However the subtraction of the ZPF
suggests the following rhetorical question: How the Alice detector knows
that it should act according to the field intensity $\left| A_{1}\right|
^{2} $ when the actual arriving intensity is $\left| A\right| ^{2}?$ (see eq.%
$\left( \ref{3.3}\right) $ for the notation)$.$ Or alternatively, how the
detector knows what intensity would arrive when the pumping laser was
switched off in order to subtract it?. My answer is as follows. We must
assume that the intensity arriving at any point of space when there is no
external action (i.e. for free EM field) is always given by eq.$\left( \ref
{W0}\right) ,$ in particular the radiation arriving at the said point is
isotropic. Then it is plausible to assume that only deviations from
isotropy, due to signals, could excite a detector. It is similar to the
effect of air on bodies, that produces relevant consequences when the
pressure of wind on the body is anisotropic. The intensity, $\left|
A_{0}\right| ^{2}$ , that would arrive at Alice coming from the source (the
nonlinear crystal) when the laser is switched off, is a part of the ZPF.
Therefore it would be balanced by a similar intensity coming from all other
directions. In sharp contrast $\left| A_{1}\right| ^{2}$ is just the clean
intensity arriving at the detector, which was originated by the action of
the laser and superposed to the ZPF.

In respect to the coincidence detection, the choice eq.$\left( \ref{5.0}%
\right) $ is plausible because only the part of the radiation arriving at
Alice which is correlated with the radiation arriving at Bob should
contribute to the coincidence detection. And similarly for the radiation
arriving to Bob.

In spite of these arguments the possibility of a local realistic
interpretation of the (claimed loophole-free) violations of Bell
inequalities is not obvious. In fact our study in section 4 involves
hypotheses that, although assumed plausible, might be flawed. There is also
a contradiction which should be explained. The derivation of Bell
inequalities with reference to the past light cone of the detection events 
\cite{Bell} seems to be valid in the case analyzed in section 4. Even if we
assume that the ZPF is real, the field arriving at the detectors can be
influence just be events in the past light cone and any possible action of
the measurement of Alice (Bob) of the detection by Bob (Alice) is carefully
excluded in the performed experiments. This contradiction between our
results eqs.$\left( \ref{3.5}\right) $ and $\left( \ref{5.4}\right) $ and
the quite general derivation of the Bell inequalities \cite{Bell} is worth
to be studied. I believe that a hint for the solution is the fact that
detection of electromagnetic waves cannot been assumed \textit{an event}. It
is a process with a duration much larger than the periods of the involved
fields. This question will not be studied further in the present paper, but
it certainly deserves future research.

\end{document}